\documentclass[a4paper,11pt]{article}
\usepackage{jheppub}
\usepackage{overpic}

\pdfoutput=1
\usepackage{CJK}
\usepackage{xcolor}
\usepackage{amsfonts}
\usepackage{epstopdf}
\usepackage{wrapfig} % to wrap figure with
\usepackage{subfigure}
\usepackage{graphicx}  % Include figure files
\usepackage{dcolumn}   % Align table columns
\usepackage{bm}
\usepackage{float}
\usepackage{lipsum}

\usepackage{epsfig}
\usepackage{psfrag}
\usepackage{amsmath}
\usepackage{amssymb}
\usepackage{color}
\usepackage{hyperref}
\hypersetup{colorlinks,linkcolor={blue},citecolor={blue},urlcolor={blue}}

\newcommand{\be}{\begin{equation}}
\newcommand{\ee}{\end{equation}}
 \newcommand{\bea}{\begin{eqnarray}}
 \newcommand{\ena}{\end{eqnarray}}

\begin{document}

\newcommand\blfootnote[1]{%
\begingroup
\renewcommand\thefootnote{}\footnote{#1}%
\addtocounter{footnote}{-1}%
\endgroup
} %footnote on front page

\setlength{\parskip}{0.3em}
\title{Topological Defects Formation with Momentum Dissipation}

% more complex case: 4 authors, 3 institutions, 2 footnotes
\author[1]{Zhi-Hong Li,}
\author[2]{Hua-Bi Zeng,}
\author[1,3]{Hai-Qing Zhang}

\affiliation[1]{Center for Gravitational Physics, Department of Space Science, Beihang University, Beijing 100191, China}
\affiliation[2]{Center for Gravitation and Cosmology, College of Physics Science and Technology, Yangzhou University, Yangzhou 225009, China}
\affiliation[3]{International Research Institute for Multidisciplinary Science, Beihang University, Beijing 100191, China}

% e-mail addresses: one for each author, in the same order as the authors
\emailAdd{lizhihong@buaa.edu.cn}
\emailAdd{hbzeng@yzu.edu.cn}
\emailAdd{hqzhang@buaa.edu.cn}

%\begin{abstract}
\abstract{
We employ holographic techniques to explore the effects of momentum dissipation on the formation of topological defects during the critical dynamics of a strongly coupled superconductor after a linear quench of temperature. The gravity dual is the dRGT massive gravity in which the conservation of momentum in the boundary field theory is broken by the presence of a bulk graviton mass. From the scaling relations of defects number and ``freeze-out" time to the quench rate for various graviton masses, we demonstrate that the momentum dissipation induced by graviton mass has little effect on the scaling laws compared to the Kibble-Zurek mechanism. Inspired from Pippard's formula in condensed matter, we propose an analytic relation between the coherence length and the graviton mass, which agrees well with the numerical results from the quasi-normal modes analysis. As a result, the coherence length decreases with respect to the graviton mass, which indicates that the momentum dissipation will augment the number of topological defects.
}

\maketitle
%\pagebreak

\section{Introduction}
\label{sec:intro}

Non-equilibrium dynamics of strongly coupled system is one of the most challenging tasks in theoretical physics. It is not only interesting in cosmology and particle physics, but also interesting within condensed matter physics \cite{henkel,hohenberg,Polkovnikov:2010yn}. In recent years, the central question in understanding the strongly coupled theory was how the critical dynamics across a phase transition react to a time dependent coupling, i.e., a thermal or quantum quench.  We investigate the critical dynamics of a strongly coupled superconducting phase transition within the AdS/CFT correspondence from string theory \cite{Maldacena:1997re}, focusing on the formation and evolution of topological defects after a thermal quench. Spontaneous topological defect formation in such transition is a fundamental component of the celebrated Kibble-Zurek mechanism (KZM) \cite{Kibble:1976sj,Kibble:1980mv,Zurek:1985qw}.

The basic idea of KZM is that: As a continuous symmetry is spontaneously broken during a second order phase transition, the evolution of the system stops being adiabatic and enters the ``freeze-out'' region as a result of the critical slowing down nearby the critical point. Topological defects will form at the interfaces where different symmetry-breaking domains meet and satisfy the ``geodesic rule'' \cite{Bowick:1992rz}. KZM predicts a universal power law between the topological defects number and the quench rate. Such mechanism was first introduced by Kibble \cite{Kibble:1976sj,Kibble:1980mv} into Cosmology that as a result of relativistic causality, topological defects such as cosmic strings, monopoles, and etc., can form in the subsequent evolution of the Universe. It was later introduced by Zurek into condensed matter physics that the vortex lines may occur in a superfluid when the system is quenched, which is accessible in the laboratory \cite{Zurek:1985qw}. %The basic idea of KZM is that when a system is quenched through a critical point of a continuous second-order phase transition, where the system slows down and at certain temperature different domains in the system cannot communicate with each other. The critical slowing down means that new broken symmetry phase is chosen and the choice must be made locally, within regions that the dynamics essentially freezes, and there is a breakdown of adiabaticity. And leading to the formation of topological defects at the vertex of different domains.
 This scenario was supported by various numerical studies \cite{Laguna:1996pv,Yates:1998kx,Ibaceta:1998yy,Antunes:1998rz,Donaire:2004gp,Das:2011cx,Gillman:2017ycq}, and experiments carried out in a variety of systems \cite{Chuang:1991zz,Bowick:1992rz,Digal:1998ak,Baeuerle:1996zz,Ruutu:1995qz,Carmi:2000zz,Monaco:2002zz,Maniv:2003zz,Golubchik:2010zz,Ado1,Ado2}.

% Gauge/gravity duality (AdS/CFT correspondence) is a ``first-principle'' means to study the strongly coupled field theories from weakly coupled gravitational theories in one higher dimensions \cite{Maldacena:1997re}.  Previous holographic studies on KZM can be found in \cite{Chesler:2014gya,Sonner:2014tca,Zeng:2019yhi}.
%{\color{red}{How to connect}}

The theory of KZM was proposed in a spatially homogenous background in which the momentum is conserved. However, real materials, which are composed of electrons, atom lattices, impurities and so on, do not possess such kind of momentum conservations (or have conservation only modulo the reciprocal lattice vectors) due to the spatial inhomogeneities \cite{MT}. It will be quite important to investigate the KZM in the strongly coupled systems with momentum dissipation in the framework of holography.  From holography, there have been a number of models built to study momentum dissipation by implementing the effects of translational symmetry breaking. Several approaches among these are to include the lattice effects \cite{Hartnoll:2012rj},  or study the charge transport in the background of a spatially modulated bulk \cite{Horowitz:2012ky,Horowitz:2012gs,Horowitz:2013jaa,Donos:2012js}, or in the presence of impurities  directly \cite{Hartnoll:2008hs,Anantua:2012nj}. However, due to the involved technics in such models, another conceptually distinct approach to holographic momentum dissipation was suggested in \cite{Vegh:2013sk}. This approach does not make use of any conventional mechanism of translational symmetry breaking, but instead provides an effective bulk description of a theory without momentum conservation from massive gravity. Applications of gauge/gravity duality on non-equilibrium dynamics can be found in \cite{Liu:2018crr,Guo:2018mip,Gao:2019baf,Lan:2020kwn,Wittmer:2020mnm,Ewerz:2020wyp,Erdmenger:2020flu,Ecker:2021ukv}.
Previous holographic work on KZM were carried out in \cite{Sonner:2014tca,Chesler:2014gya,Zeng:2019yhi,Li:2019oyz,delCampo:2021rak,Xia:2020cjl,Li:2021dwp}, without any momentum dissipation in the background.

Holographic massive gravity exhibits momentum dissipation \cite{Davison:2013jba} by giving a mass to the graviton. This mass breaks the diffeomorphism invariance of the gravitation theory, via the holographic dictionary, it in turn violates the conservation of energy-momentum in the dual field theory. Many clues have shown that the effect of massive graviton in the bulk gravity can be considered as the effect from the lattice in the dual field theory on the boundary \cite{Blake:2013owa,Hu:2015dnl}. Recently, a nonlinear massive gravity theory has been proposed by de Rham et.al. (dRGT theory) \cite{deRham:2010ik,deRham:2010kj}, and later it is found to be ghost-free \cite{Hassan:2011hr,Hassan:2011tf}. Note that, the dRGT massive gravity is thought to be the only healthy theory in Poincar\'{e} invariant setups, and there have been numerous studies in dRGT massive gravity which we will not list all of them \cite{Blake:2013bqa,Cai:2014znn,Hu:2015xva,Xu:2015rfa,Hendi:2015hoa}.

Our goal in this paper is to explore the effect of the momentum dissipation (equivalently the effect of graviton mass) on the critical dynamics across a second-order phase transition after a linear quench of temperature from the holographic $(3+1)$-dimensional dRGT massive gravity. Quantized superconductor vortices will turn out due to KZM after quench. We will see that the momentum dissipation has little effect on the KZM scaling laws of the vortex number and ``freeze-out'' time to the quench rate. However, it is interesting to see that the coherence length $\hat\xi$ of the order parameter at the ``freeze-out'' time decreases with respect to the graviton mass $m$ from the quasi-normal modes (QNMs) analysis. Physically, it implies that the momentum dissipation will reduce the coherence length. Inspired from Pippard's formula for the coherence length and mean free path \cite{abp}, we propose an analytic expression between the coherence length $\hat\xi$ and the graviton mass $m$. This analytic relation is in good agreement with the results from QNMs analysis. Moreover, we also investigate the relations between the number of vortices $n$ and the graviton mass $m$ from both numerical simulations and analytic approach, and find that they fit very well as $m$ is not too large. The resulting vortex number will increase as graviton mass grows, which in turn indicates that momentum dissipation will enlarge the number of ovrtices.

%From these studies, we conclude that the graviton mass (or equivalently momentum dissipation) is the main factor affecting the correlation length $\xi$ in massive gravity and verify the relationship between vortex number density and correlation length $n \varpropto \xi^{-2}$.

 %relation between the graviton mass $m$ and the coherence length $\xi$  from QNMs, and we find the coherence length decreases with respect to the graviton mass.   we investigate the relation of the vortex number density , and our numerical results show that the graviton mass parameter in dRGT massive gravity can increase the number of vortex.

 The outline of the paper is as follows. In Sec.\ref{KZM}, we briefly review the KZM, and build up the holographic model in the (3+1)-dimensional dRGT massive gravity in in Sec.\ref{sec:setup}; Sec.\ref{numres} contains the main analytical and numerical results of our paper to study the generations of vortices and the relations between graviton mass;  The conclusions are drawn in Sec.\ref{sec:conclusion}; In the Appendix \ref{app} we briefly introduce the dRGT massive gravity.

 %{\red at the microscopic level, momentum dissipation could be achieved by breaking the translational symmetry of the system with a lattice or impurities, and this would also invalidate the constitutive relations of the theory. from https://arxiv.org/pdf/1411.1062.pdf}

\section{A brief review of Kibble-Zurek mechanism}
\label{KZM}
KZM is a paradigmatic theory to describe spontaneous generation of topological defects  in non-equilibrium critical dynamics \cite{Kibble:1976sj,Kibble:1980mv,Zurek:1985qw}. In the vicinity of critical point of a second-order phase transition, critical opalescence and critical slowing-down implies that the coherence length $\xi$ and the relaxation time $\tau$ diverge as
\begin{alignat}{1}\label{eq:9}
 \xi(\epsilon)=\xi_0|\epsilon|^{-\nu}, \qquad  \tau(\epsilon)=\tau_0|\epsilon|^{-\nu z},
\end{alignat}
in which $\xi_0$ and $\tau_0$ are constant coefficients depending on the microscopic physics, while $\nu$ and $z$ are the static and dynamic critical exponents in equilibrium states, and $\epsilon$ is the dimensionless distance to the critical temperature, $\epsilon\equiv1-T/T_c$. KZM assumes a linear quench across the critical point as $\epsilon(t)=t/\tau_Q$, in which $\tau_Q$ is the quench rate (or quench time). The ``freeze-out" time occurs at the instant $\hat t$ when the relaxation time of the system $\tau$ equals the time scale of the quench to the critical point, then the dynamics approximately freezes (from nearly adiabatic to approximately impulse behavior within the time interval [$-\hat t$, $\hat t$]).  Therefore, the ``freeze-out'' time can be readily obtained as
\begin{eqnarray}\label{eq:tfreeze}
\hat t=(\tau_0\tau_Q^{\nu z})^{\frac{1}{1+\nu z}}.
\end{eqnarray}
The key insight of the KZM is that the average size $\hat\xi$ of the domains in the broken symmetry phase is set by the the equilibrium coherence length $\xi$ evaluated at the freeze-out time $\hat t$, i.e., $\hat\xi=\xi(\hat t)$. Thus, the size of the symmetry-breaking domain is given by,
\begin{eqnarray}
  \hat\xi=\xi_0\left(\frac{\tau_Q}{\tau_0}\right)^{{\nu}/{(1+\nu z)}}
\end{eqnarray}
Topological defects will occur at the interfaces between various symmetry-breaking domains if they satisfy the ``geodesic rules'' \cite{Bowick:1992rz}.  Therefore, the above estimate of the $\hat\xi$ can be recast as an estimate for the resulting number of topological defects,
\begin{alignat}{1}\label{density}
n\approx \frac{Ap}{\hat\xi^{d}}=\frac{Ap}{\xi_0^{d}}\left(\frac{\tau_0}{\tau_Q}\right)^{\frac{d\nu}{1+\nu z}}.
\end{alignat}
where $A$ is the volume size of the system, $p$ is the average probability for the domains to form defects and $d$ is the spatial dimension of the system. This relation is verified to be universal in various dimensions and for different kinds of topological defects \cite{Laguna:1996pv,Yates:1998kx,Ibaceta:1998yy,Antunes:1998rz,Donaire:2004gp,Das:2011cx,Gillman:2017ycq,Chuang:1991zz,Bowick:1992rz,Digal:1998ak,Baeuerle:1996zz,Ruutu:1995qz,Carmi:2000zz,Monaco:2002zz,Maniv:2003zz,Golubchik:2010zz}.  In particular, the generated topological defects are vortices for superfluids or superconductors in $d=2$ system. We will investigate the KZM for vortices in a strongly coupled superconductor system with background momentum dissipation by virtue of the AdS/CFT technique in the following.

%It claim that the KZM holds and show that the non-equilibrium dynamics across the phase transition is also universal. This requires the ability to measure the average number of excitations after driving the system at a given quench rate, and repeating this measurement for different quench rates.

\section{Holographic setup}
\label{sec:setup}
The gravity background we adopt is the dRGT massive gravity in $(3+1)$-dimensional bulk spacetime \cite{deRham:2010ik,deRham:2010kj} (Please consult Appendix \ref{app} for a brief review of dRGT massive gravity). In order to study the dynamics in this background, we work under the Vaidya-AdS solution with flat space $k=0$ and zero charge $q=0$ (We have scaled the AdS radius $L\equiv1$.).
\begin{equation}\label{eq:1}
ds^{2}=\frac{1}{z^{2}}\left[-f(z)dt^2-2dt dz+\left(dx^{2}+dy^{2}\right)\right],
\end{equation}
where
\begin{equation}
f(z)=1-M z^3+\frac{c_1m^2}{2}z+c_2m^2z^2,
\end{equation}
in which $M$ is the mass parameter of the black hole, $c_i~(i=1,2)$ are constants and $m$ stands for the graviton mass.
%\footnote{{\red To be precise, $m$ is the graviton mass near the UV boundary. We know the effective graviton mass depend on the radial direction from \cite{Blake:2013bqa}. Here $m$ is proportional to the graviton mass near the UV boundary up to a constant by taking the radial direction to the UV boundary. We call $m$ here as graviton mass for simplicity.}}
Location of the AdS boundary is $z=0$ while the horizon is $z=z_+$. Hawking temperature of the black brane thus is
\begin{eqnarray}
T=\frac{1}{4\pi}\left(\frac{3}{z_+}+{c_1m^2}+{c_2m^2}z_+\right).
\end{eqnarray}
We work in the probe limit and adopt the conventional Maxwell-complex scalar action for holographic superconductors \cite{hartnoll}
\begin{eqnarray}\label{eq:3}
  S_{\rm matter} =\int d^4 x\sqrt{-g} \left[-\frac{1}{4}F_{\mu\nu}F^{\mu\nu}-|D_\mu\Psi|^2-m_\psi^2|\Psi|^2\right]
\end{eqnarray}
in which $F_{\mu\nu}=\partial_{\mu}A_\nu-\partial_{\nu}A_\mu$ is the Maxwell field strength, $A_\mu$ is the $U(1)$ gauge field, $D_\mu=\nabla_\mu-iA_\mu$ is the covariant derivative and $m_\psi$ is the mass of the complex scalar field $\Psi$. Thus, the equations of motion can be obtained
%by ignoring the backreaction of the matter fields to the gravitational fields. The EoMs can be obtained readily from the above action as
\begin{eqnarray}
\label{EoM1}
0&=&(\nabla_\mu-iA_\mu)(\nabla^\mu-iA^\mu)\Psi-m_\Psi^2\Psi,\\
\nabla_\nu F^{\nu\mu}&=&i(\Psi^\ast(\nabla^\mu-iA^\mu)\Psi-\Psi(\nabla^\mu+iA^\mu)\Psi^\ast).\label{EoM2}
\end{eqnarray}
It is convenient to work in axial gauge by setting $A_{z}=0$, and take the ansatz for other fields as $\Psi=\Psi(t,z,x,y),A_{t,x,y}=A_{t,x,y}(t,z,x,y)$.

\subsection{Holographic renormalization and boundary conditions}
In order to solve the above equations, we need to impose suitable boundary conditions at the horizon $(z=z_+)$ and at the AdS boundary $(z=0)$.
Near the boundary $z\to0$, fields can be expanded as,
\begin{eqnarray}
  \Psi = \Psi_{0}z^{\Delta_{-}} + \Psi_{1}z^{\Delta_{+}} +\mathcal{O}(z^{\Delta_{+}+1}) ,~~~
  A_{\mu}= a_\mu +b_\mu z+\mathcal{O}(z^2) ,
\end{eqnarray}
with $\Delta_{\pm}={\left(3\pm\sqrt{9+4m_\psi^2}\right)}/{2}$ are the conformal dimensions of dual scalar operators on boundary. Following the AdS/CFT dictionary, the interpretation of $\Psi_{0}$ is that it sources the symmetry breaking operator $\hat{O}$, while $a_t$ and $a_i$ are the chemical potential and superfluid velocity in the boundary, respectively. The physical meaning of $\Psi_1$ and $b_\mu$ will be clear after the holographic renormalization.

From holographic renormalization \cite{Skenderis:2002wp}, varying the renormalized on-shell action with respect to the source terms one can achieve the corresponding conjugate variables.
In order to obtain finite on-shell action, we should add the counter terms of the scalar fields $S_{\rm ct}=\int d^3x\sqrt{-\gamma}\Psi^*\Psi$ into the divergent on-shell action, where $\gamma$ is determinant of the reduced metric on the boundary. We set Neumann boundary conditions for gauge fields in order to get the dynamical gauge fields in the boundary. In order to have a well-defined variation, the surface term $S_{\rm sf}=\int d^3x\sqrt{-\gamma}n^\mu F_{\mu\nu}A^\nu$  should also be added, where $n^\mu$ is the normal vector perpendicular to the boundary. Hence, we can get the finite renormalized on-shell action $S_{\rm ren}$. The expectation value of the order parameter can be obtained from the holographic renormalization by varying $S_{\rm ren}$ with respect to $\Psi_0$, resulting in $\langle\hat{O}\rangle=\Psi_1$.

We impose $\Psi_0=0$ all time in order to have a spontaneous symmetry breaking of the order parameter. This translates into a homogenous Dirichlet boundary condition for the field $\Psi$ in the AdS boundary, i.e., $\Psi|_{z=0}=0$. Expanding the $z$-component of the Maxwell equations near boundary, we reach $\partial_tb_t+\partial_iJ^i=0$, which is exactly a conservation equation of the charge density $\rho$ and current $J^i$ on the boundary, since from the variation of $S_{\rm ren}$ one can get $b_t=-\rho$ and $J^i=-b_i-(\partial_ia_t-\partial_ta_i)$. At the horizon, we demand the regularity of the fields. Since the metric component $g_{tt}$ is zero at the horizon, the component $A_t$ should be vanishing, while other fields are finite at the horizon.

\subsection{Numerical schemes}
\begin{figure}[h]
\centering
\includegraphics[trim=3.5cm 9.5cm 3.5cm 9.5cm, clip=true, scale=0.6,  angle=0] {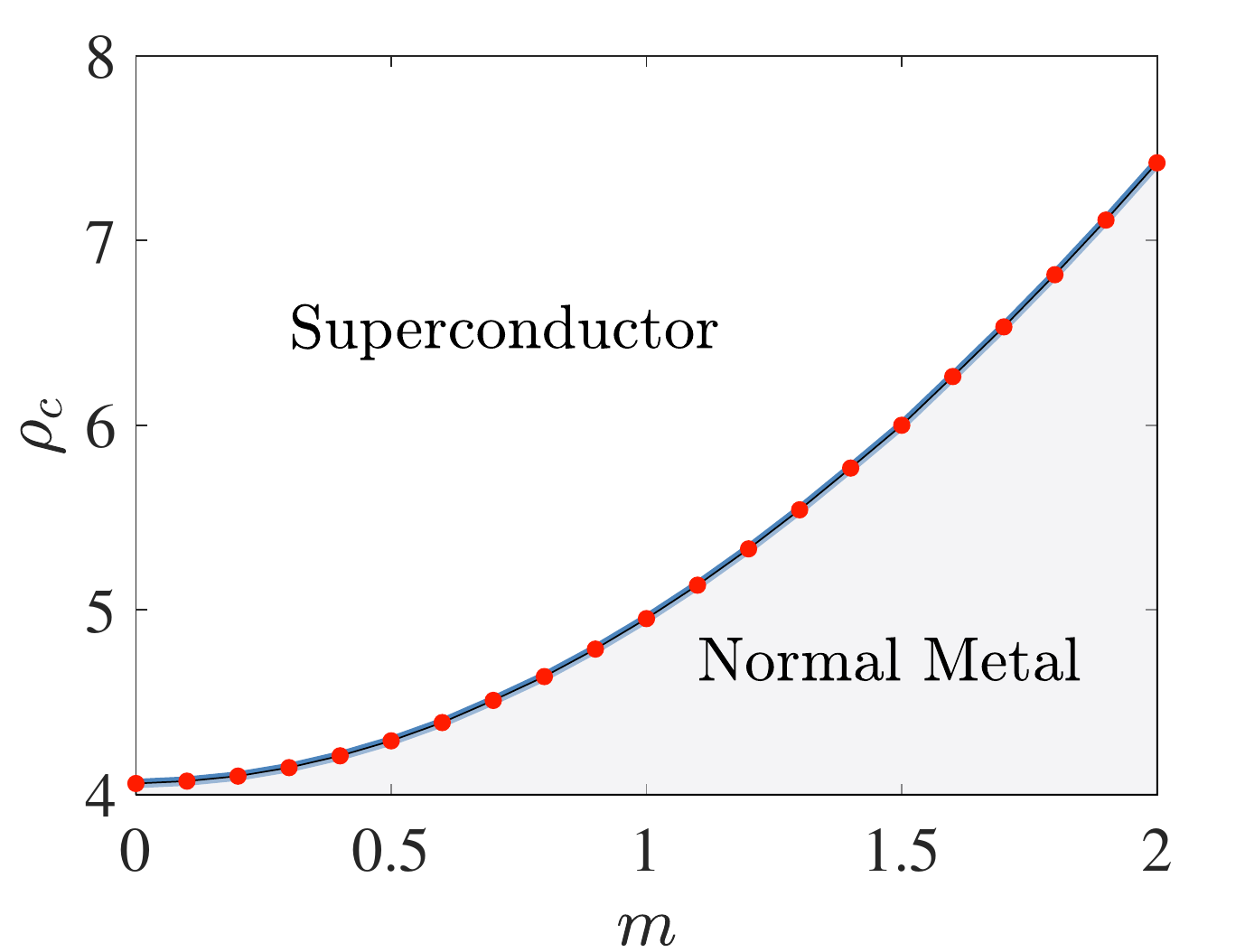}
\caption{Critical charge density versus graviton mass. Red dots are numerical results while the blue line is a fitted critical curve. Upper the critical curve the system is in a superconducting phase while below the curve it is in a normal metal phase (shaded region). }\label{rhom}
\end{figure}

We choose $m_\psi^2=-2$ without loss of generality and set $c_1=1, c_2=-1/2$ in order to make the background thermodynamically stable \cite{Hu:2015dnl,Cai:2014znn}. For convenience of the numerical calculation, we set the black hole mass parameter as $M=1$ and the location of horizon at $z_+=1$. From holographic superconductor \cite{hartnoll}, increasing the chemical potential or the charge density equals decreasing the temperature of the system. We choose to increase the charge density $\rho$ in our paper. In addition, the temperature of the black hole $T$ has the mass dimension one while the charge density $\rho$ has mass dimension two from the dimensional analysis. Therefore, in order to linearize the temperature near the critical point according to KZM, we quench the charge density $\rho$ as $\rho(t)=\rho_c(1-t/\tau_Q)^{-2}$, where $\rho_c$ is the critical charge density for the static and homogeneous holographic superconducting system. We should note that different graviton masses $m$ correspond to different $\rho_c$, which is shown in Fig.\ref{rhom}

We quench the system from normal state with initial temperature $T_i=1.4T_c$ to superconducting state with the final equilibrium temperature $T_f=0.8T_c$. The size of the boundary system is $(x,y)=(50,50)$. We take advantage of the Chebyshev pseudo-spectral method in the radial direction $z$ and use the Fourier decomposition in the $(x, y)$-directions since the periodic boundary condition is imposed along $(x, y)$-directions. The grids in $z$-direction is 21 while the grids along $(x, y)$-directions are $201\times201$.  We thermalize the system by adding random seeds for the fields at the initial time.  The random seeds are sampled in the bulk by satisfying the white noise distributions $\langle s(t, \vec x)\rangle=0$ and $\langle s(t, \vec x)s(t',\vec x')\rangle=\zeta\delta(t-t')\delta(\vec x-\vec x')$, with the amplitude $\zeta\approx10^{-3}$. The system evolves by using the 4th order Runge-Kutta method with time step $\Delta t=0.01$. Filtering of the high momentum modes are implemented following the ``2/3's rule'' that the uppermost one third Fourier modes are removed \cite{Chesler:2013lia}.

%We thermalize the system by adding small random seeds in the normal state before quenching.

%{\color{blue}{We take advantage of the Chebyshev pseudo-spectral method with 21 grids in the radial direction $u$ and use the Fourier decomposition in the $(x,y)$-directions since the periodic boundary condition along $(x,y)$ was imposed. We thermalize the system by adding small random seeds in the normal state before quench. The reason is to make sure that the system before quench is in a symmetrical phase, which is the requirement of KZM.  Different from putting the seeds on the boundary in \cite{Chesler:2014gya,Sonner:2014tca}, we add the random seeds of the fields in the bulk by satisfying the distributions $\langle s(t,x_\mu)\rangle=0$ and $\langle s(t,x_\mu)s(t',x'_\mu)\rangle=\zeta\delta(t-t')\delta(x_\mu-x'_\mu)$ where $(\mu=z,x,y)$, with the amplitude  $\zeta\approx10^{-3}$. \footnote{Other relatively smaller magnitudes of $\zeta$ lead to similar results. In principle, $\zeta$ cannot be too large since the seeds play the role of perturbations to thermalize the system.}  The system evolves by using the fourth order Runge-Kutta method with time step $\Delta t=0.07$ for small value of graviton mass and $\Delta t=0.01$ for larger value of graviton mass. Filtering of the high momentum modes are implemented following the ``$2/3$'s rule'' that the uppermost one third Fourier modes are removed \cite{Chesler:2013lia}.}}

%\section{Quasi-normal modes of the scalar field}

\section{Results}
\label{numres}
\subsection{KZM scalings of vortex number and ``freeze-out" time}
\label{sec:density}
\begin{figure}[h]
\centering
\includegraphics[trim=3.2cm 9.2cm 4.3cm 10cm, clip=true, scale=0.55,  angle=0] {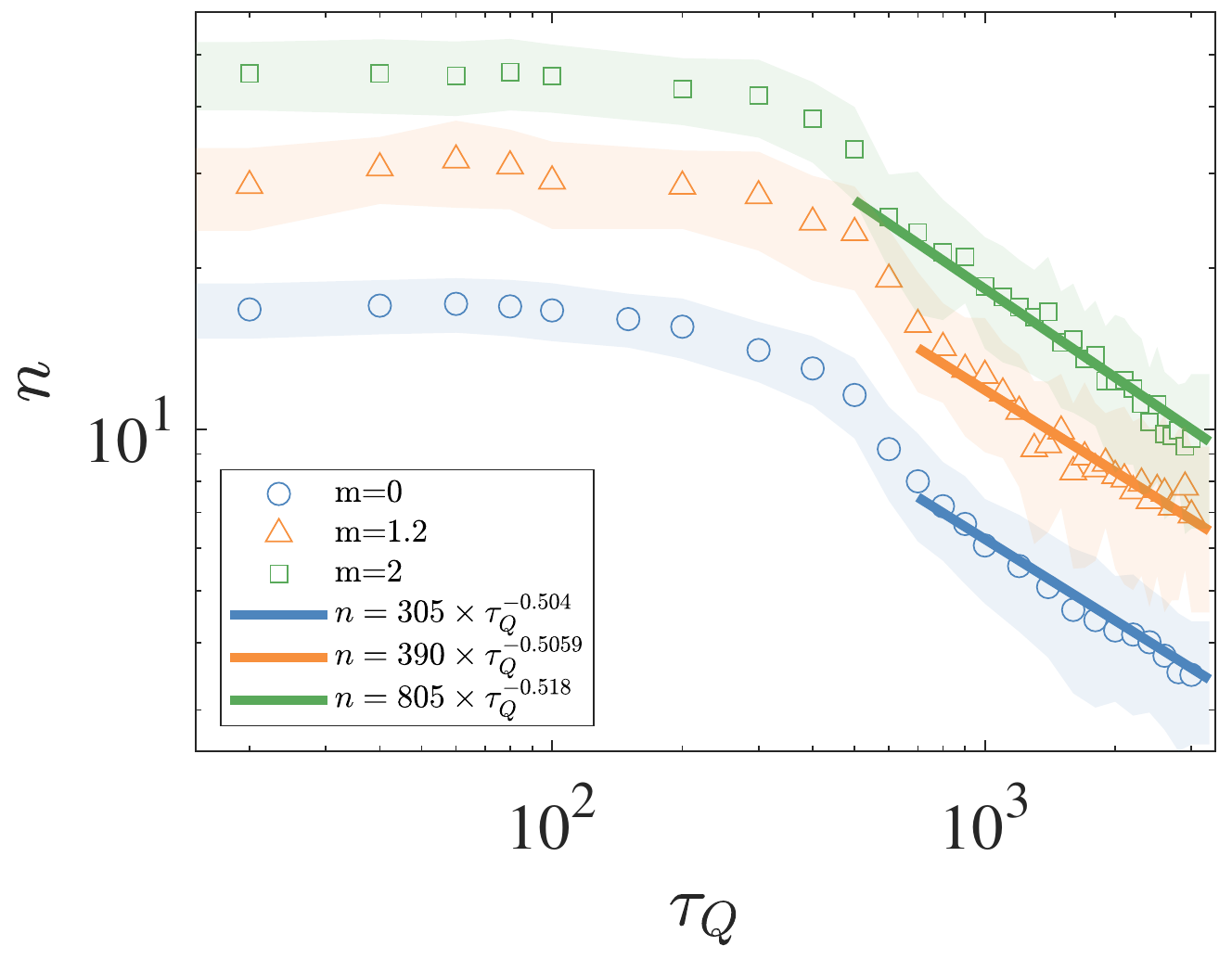}~~%{TqN.pdf}~
\includegraphics[trim=3.2cm 9.2cm 4.3cm 10cm, clip=true, scale=0.55,  angle=0] {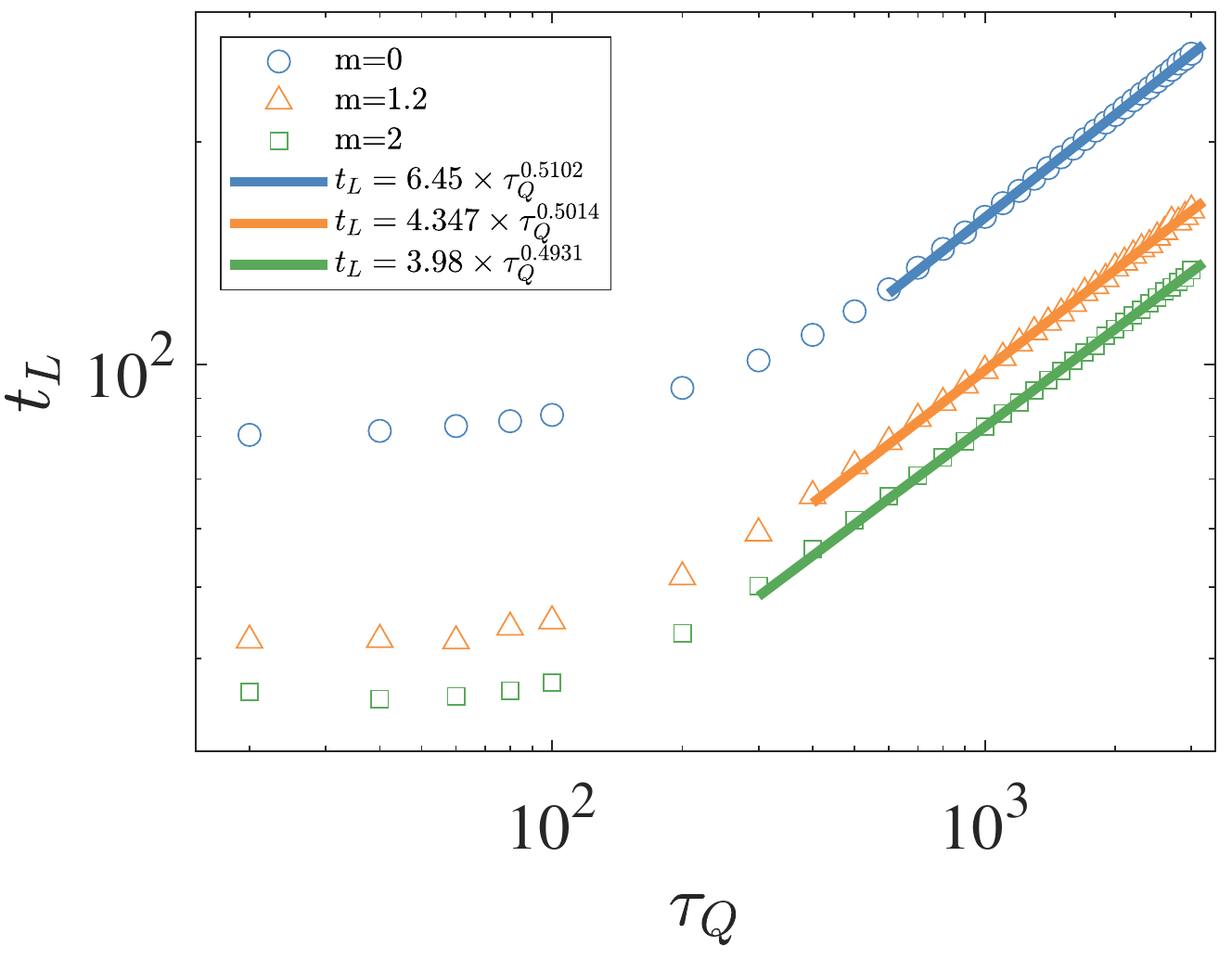}%{Tqtlag.pdf}
\caption{\label{NT}
 (Left panel) Double logarithmic relation between vortex number $n$ and quench time $\tau_Q$ for graviton mass $m=(0, 1.2,2)$. The circles, triangles and squares are the numerical data while the straight lines are the best fitted lines in the slow quench regime (larger $\tau_Q$). The shaded regions represent the standard deviations away from the mean values.  For the slow quench, the scaling relations satisfy the KZM scaling laws very well; However, for fast quench (smaller $\tau_Q$), vortex number $n$ is almost constant beyond the scope of KZM. (Right panel) Double logarithmic relation between the lag time $t_L$ and quench rate $\tau_Q$. For slow quench the relation between $t_L$ and $\tau_Q$ satisfy the KZM scalings very well, while for fast quench $t_L$ keeps almost constant. This figure was obtained by running $100$ times of simulations for each $\tau_Q$ and $m$. }
\end{figure}

In this subsection, we will examine the KZM scaling laws in Eqs.\eqref{eq:tfreeze} and \eqref{density}, depending on the graviton mass. We count the vortex number $n$ as the order parameter $\langle\hat O\rangle$ saturates to its equilibrium value. Scalings between the number $n$ and quench time $\tau_Q$ under different graviton mass $m$ are exhibited in the left panel of Fig.\ref{NT}. The shaded regions correspond to the standard deviations from the mean values.  In the fast quench regime (small $\tau_Q$) the vortex number approximately saturates due to the finite size effect, which is consistent with previous results in condensed matter or holography \cite{Chesler:2014gya,Sonner:2014tca,Zeng:2019yhi,Li:2019oyz}.  However, in the slow quench regime (large $\tau_Q$), the vortex number will decrease with respect to quench time satisfying the scaling relations as $n_{m=0}\approx 305\times\tau_{Q}^{-0.504}$, $n_{m=1.2}\approx 390\times\tau_{Q}^{-0.5059}$ and $n_{m=2}\approx 805\times\tau_{Q}^{-0.518}$ respectively. From this figure, we can conclude that as $m$ grows, or equivalently momentum dissipation gets stronger, the number of vortices will increase as well. In other words, from the relation $n\propto 1/\hat \xi^2$ we can deduce that the momentum dissipation will reduce the coherence length. We will study this relation between coherence length and the graviton mass in the next subsection in details.

%for slower quench, the scaling laws between $n$ and $\tau_Q$ for $z=1$ is roughly $n =n_1 \tau_Q^a$ with $n_1\approx(703.9170\pm1.1333)$ and $a\approx (-0.4998\pm0.0162)$, in which the error bars stand for the standard deviations. For $m=(1.2,2.5)$, the relation are $n\approx (482.62\pm1.1431)\times\tau_{Q}^{(-0.4897\pm0.0183)}$ and $n\approx (312.9211\pm1.2379)\times\tau_{Q}^{(-0.4961\pm0.0315)}$ respectively.
%In the left panel of Fig.\ref{NT}, we show the relation between the number density of vortices $n$ and quench time $\tau_Q$ under different graviton mass $m$.

The ``freeze-out'' time $\hat t$ can be reflected by studying the lag time $t_L$ that defined as order parameter begins to grow rapidly \cite{Das:2011cx,Sonner:2014tca,Zeng:2019yhi}. In numerics we operationally set $t_L$ as $\langle \hat O\rangle\sim0.1$ following \cite{Sonner:2014tca,Zeng:2019yhi,Das:2011cx}.
On the right panel of Fig.\ref{NT} we exhibit the relation between $t_L$ and $\tau_Q$. The error bars are not shown since they are very tiny. We see that for fast quench the lag time is almost constant for different $m$'s. However, for slow quench, one can read that $t_{L}\approx6.45\times\tau_{Q}^{0.5102}$ for $m=0$, $t_{L}\approx4.347\times\tau_{Q}^{0.5014}$ for $m=1.2$ and $t_{L}\approx3.98\times\tau_{Q}^{0.4931}$ for $m=2$.
 Therefore, from the two KZM scaling relations in Eqs.\eqref{eq:tfreeze} and \eqref{density}, one can readily evaluate the dynamic critical exponent $z$ and the static critical exponent $\nu$ on the boundary as $(\nu\approx 0.514496, z \approx 2.0246), (\nu \approx 0.50732, z \approx 1.98221)$ and $(\nu \approx 0.510949, z \approx 1.90386)$ for $m=(0, 1.2, 2)$ respectively, in which the exponent matches the mean-field theory values with $z=2$ and $\nu=1/2$. This means graviton mass in dRGT massive gravity does not alter the KZM scaling laws,  which in turn indicates that the existence of graviton mass does not affect critical exponents, $z$ and $\nu$, in the boundary field theory. The boundary field theory remains as a mean-field theory, which is consistent with the assumption of AdS/CFT correspondence that the boundary is a mean-field theory in large $N$ limit \cite{Zaanen:2015oix}.

\subsection{Coherence length and graviton mass}

\subsubsection{Quasi-normal modes analysis}
From previous section \ref{KZM}, we see that the average size of the symmetry-breaking domains $\hat \xi$ is approximately the coherence length of the order parameter at the ``freeze-out'' time $\hat t$ \cite{Zurek:1985qw}.
Fortunately, from holography one can compute the coherence length from QNMs of the scalar field in the bulk by choosing suitable boundary conditions \cite{Kovtun:2005ev,Maeda:2009wv}. The QNMs correspond to the poles of the retarded Green's function of the dual field theory. One can read off the coherence length $\xi$ from the correlation function
\begin{eqnarray}\label{cof}
\langle O(\omega,k)O^\dag(-\omega,-k)\rangle\sim\frac{1}{i\tilde{c}\omega+k^2+1/\xi^2}.
\end{eqnarray}
where $k$ and $\omega$ are the momentum and frequency of perturbation modes respectively, while $\tilde{c}$ is a parameter.  QNMs are computed from small fluctuations of the background fields, such that
$\Psi(t,z,x)=\psi_0(z)+\psi(z)e^{-i\omega t+ikx}$, where $\psi_0(z)$ is the static background of scalar field while $\psi(z)e^{-i\omega t+ikx}$ is small fluctuations with respect to the background. For simplicity, we have assumed the fluctuations as plane waves in the only $x$-direction. We calculate the QNMs in the symmetric phase, i.e., in the normal phase that the background $\psi_0(z)$ is vanishing. This is because from KZM, dynamics ceases to be adiabatic and enters an impulse stage within the time interval [$-\hat t$, $\hat t$] as we learned from previous section \ref{KZM}. The starting of the symmetry breaking will occur at the instant $t\approx-\hat t$, which is in the normal phase before the critical point. This greatly simplifies the complexity of the calculations.

\begin{figure}[h]
\centering
\includegraphics[trim=3cm 9cm 4cm 9cm, clip=true, scale=0.55,  angle=0] {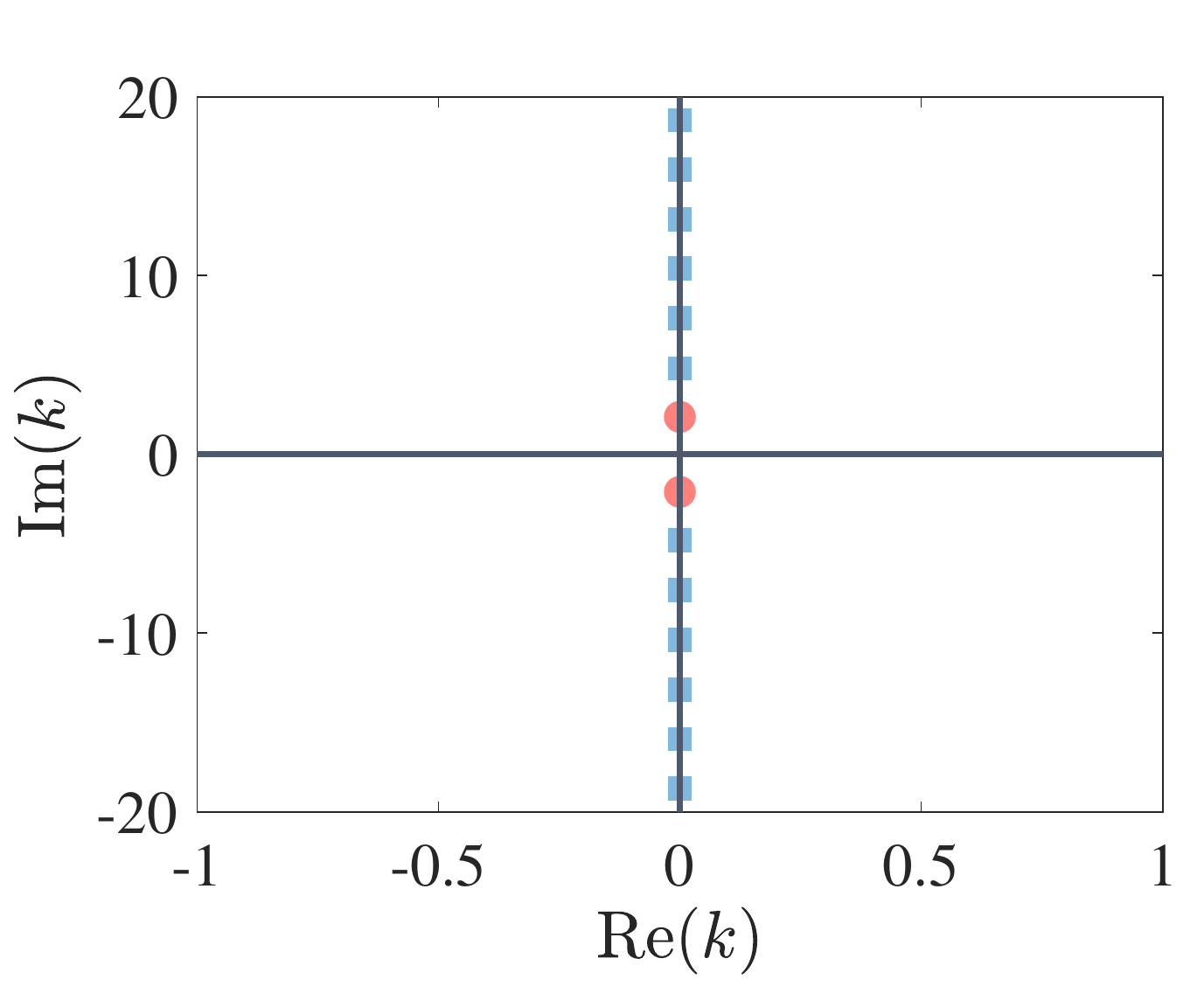}%{z2tq1800B.pdf}
\includegraphics[trim=3cm 9cm 3cm 9cm, clip=true, scale=0.55,  angle=0] {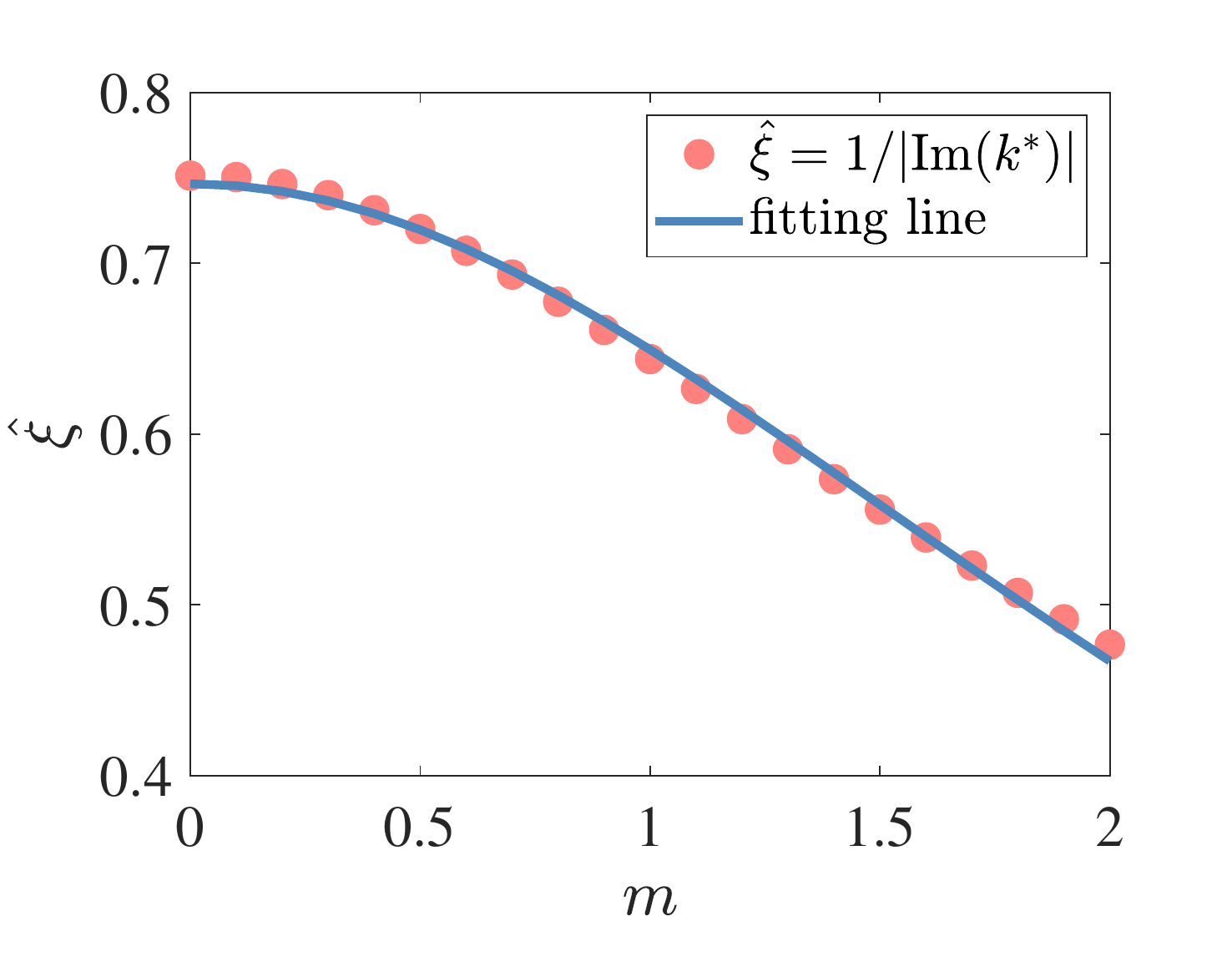}
\caption{\label{kk}{(Left panel)} QNMs in the complex $k$-plane for $m=2$ and $\omega=0$.  The leading poles nearest the real axis, marked by red dots, determine the behavior of correlations at large distances. The blue squares are the sub-leading poles which are not important to the correlations. {(Right panel)} Coherence length $\hat\xi$ of the order parameter at the ``freeze-out'' time versus the graviton mass $m$. The red dots represent $\hat\xi=1/|{\rm Im}(k^*)|$, where $k^*$ are the leading QNMs,  and the blue line is the fitting line $\hat\xi\approx0.7466/(1+0.1496m^2)$ from Eq.\eqref{eq:xim}. For both panels, we fix the quench rate as $\tau_Q=20$. }
\end{figure}

It is convenient to work out the QNMs in the Schwarzschild-AdS metric of the spacetime \cite{Sonner:2014tca,Zeng:2019yhi}.  Substituting the scalar field $\Psi(t,z,x)$ into the EoMs \eqref{EoM1},\eqref{EoM2}, and extracting the first order of the perturbations $\psi(z)$, we arrive at the decoupled linear equation as
\begin{eqnarray}\label{eq:QNM}
\psi''+\frac{f'}{f}\psi'+\left(\frac{(A_t+\omega)^2}{f^2}+\frac{zf'-k^2z^2+2}{z^2f}-\frac{2}{z^2}\right)\psi=0
\end{eqnarray}
 in which $A_t$ is the background gauge field. We impose the ingoing boundary conditions at the horizon $z_+=1$,
\begin{eqnarray}
\psi_{z_+}=(z-1)^{-\frac{iw}{4\pi T}}(1+\mathcal{O}(z-1)),
\end{eqnarray}
where $T$ is the Hawking temperature of black hole. Vanishing boundary conditions are imposed at the boundary $z=0$. We adopt the determinant method \cite{Zeng:2019yhi} to calculate the QNMs and fix the quench rate as $\tau_Q=20$ (other quench rates will have similar results). More specifically, we employ the Chebyshev differential matrix to convert the differentiation with respect to spacetime coordinate into a matrix. Zeros of the determinant of the resulting matrix are the QNMs after imposing the above boundary conditions. The lowest mode in the QNMs makes the most contribution to the perturbations. In practice, the lowest mode is the mode that is closest to the real axis. We can evaluate the coherence length from Eq.\eqref{cof} by setting $\omega=0$, and arrive at $\hat\xi=1/|{\rm Im}(k^*)|$ where $k^*$ is the lowest mode in those QNMs.  After solving the Eq.\eqref{eq:QNM} one can get a series of QNMs in the complex $k$-plane, as is shown in the left panel of Fig.\ref{kk} for $m=2$. The red dots, symmetric and closest to the real axis, stand for the lowest poles $k^*$. Other sub-leading poles  (in blue squares), all locate in the imaginary axis, are also symmetric to the real axis.   Different graviton masses correspond to different series of QNMs, leading to a relation between coherence lengths and $m$, i.e., $\hat\xi=\hat\xi(m)$, which are shown as red dots in the right panel of Fig.\ref{kk}.

\subsubsection{Analytic relation between $\hat\xi$ and $m$}

The analytic relation between the coherence length $\hat\xi$ and the graviton mass $m$ was seldom discussed. In \cite{Hu:2015dnl}, the authors found that the coherence length would decrease with respect to the graviton mass in the system of a holographic Josephson junction. However, they did not propose an analytic relation between $\hat\xi$ and $m$. In this subsection, we will seek this analytic relation between $\hat\xi$ and $m$ and show that this relation matches well with the numerical results from QNMs analysis.

From the discussions in \cite{Vegh:2013sk,Davison:2013jba,Blake:2013owa}, the graviton mass $m$ effectively plays a role of momentum dissipation in the boundary field theory. This momentum dissipation might be from scatterings of electrons to lattices or impurities, which lead to the assumption in Drude model that the relation between the mean free path $l$ and scattering time $\tau_s$ is $\tau_s=l/V_F$, in which $V_F$ is the Fermi velocity \cite{MT}.
In condensed matter physics, the coherence length $\xi$ can be related to the mean free path $l$ from Pippard's formula \cite{abp,MT},
\begin{eqnarray}\label{ppp}
\frac{1}{\xi}=\frac{1}{\xi_p}+\frac{1}{\alpha l},
\end{eqnarray}
%the interelectronic and impurity types of scattering being treated as independent.
where $\xi_p$ is the coherence length of a pure material without any scatterings, and $\alpha$ is a numerical constant.  Then from the relation $\tau_s=l/V_F$ in Drude model, we get
\begin{eqnarray}
\xi=\frac{\xi_p}{1+\frac{\xi_p}{\alpha V_F}\,\tau_s^{-1}}
\end{eqnarray}
Fortunately, from holography \cite{Blake:2013bqa} $\tau_s$ can be related to the graviton mass $m$ as $\tau_s^{-1}=\gamma m^2$ where $\gamma$ is a coefficient independent of $m$. \footnote{From the study of holographic DC conductivity in \cite{Blake:2013bqa}, the authors identified the relation between scattering time $\tau_s$ and position dependent masses $\tilde m(z)$ of perturbations along $t$-$x$ and $z$-$x$ direction at horizon as $\tau_s^{-1}=\frac{\mathcal{S}}{4\pi(\mathcal{E+P})}\tilde m^2(z_+)$. Here, $\mathcal{S}$, $\mathcal{E}$ and $\mathcal{P}$ are the entropy density, energy density and pressure of the black hole respectively. The position dependent mass at horizon is indeed proportional to the graviton mass, i.e., $\tilde m^2(z_+)\propto m^2$. We absorb all of these coefficients into $\gamma$ as $\tau_s^{-1}=\gamma m^2$ where $\gamma$ is a function independent of $m$. } Therefor, we finally reach
\begin{eqnarray}\label{eq:xim}
\xi=\frac{a}{1+b m^2}.
\end{eqnarray}
where $a=\xi_p$ and $b={\xi_p \gamma}/(\alpha V_F)$.  The fitting of this relation \eqref{eq:xim} is exhibited as blue line in the right panel of Fig.\ref{kk}, with fitting parameters as $a\approx0.7466$ and $b\approx0.1496$. From Fig.\ref{kk} we see that the numerical results (red dots) from QNMs analysis are fitted very well with the analytic relation Eq.\eqref{eq:xim}. Therefore, the coherence length $\hat\xi$ at the ``freeze-out'' time will decrease as $m$ increases. This also means stronger momentum dissipation in the boundary field will reduce the average size of the symmetry-breaking domains, which in turn implies that the vortex number $n$ would grow with respect to graviton mass according to $n\varpropto 1/\hat\xi^2$ in Eq.\eqref{density}.

\subsection{Vortex number and graviton mass}

\begin{figure}[h]
\centering
\includegraphics[trim=3.2cm 9.5cm 4.2cm 9.8cm, clip=true, scale=0.6,  angle=0] {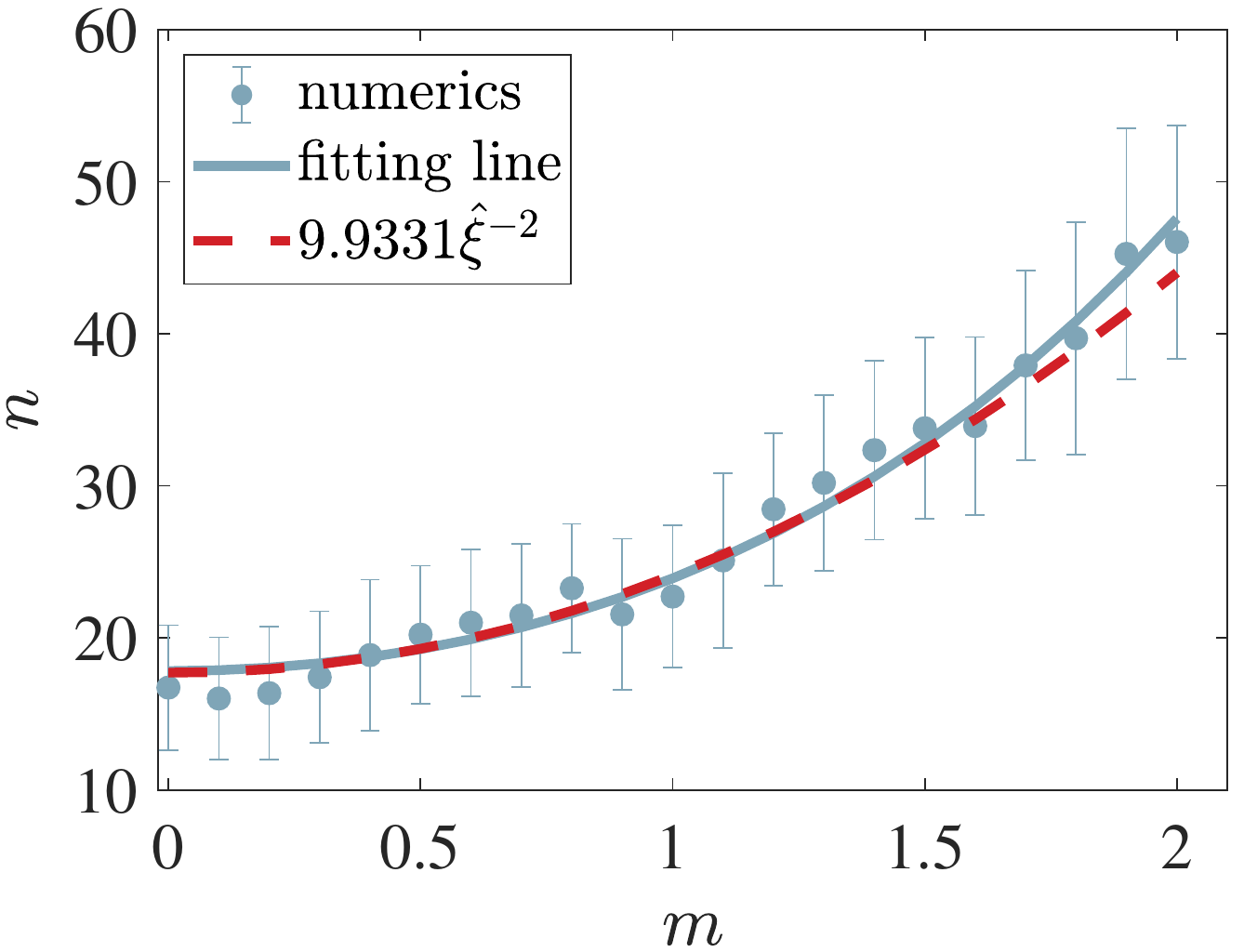}%{z2tq1800B.pdf}
\caption{\label{nm}Vortex number $n$ versus the graviton mass $m$.  Blue solid dots and error bars are from numerical data while blue line is the best fit as $n\approx17.82\times\left(1+0.1585m^2\right)^2$. Error bars denote the standard deviations. From the relation $n\approx Ap\hat\xi^{-2}$ in Eq.\eqref{density}, the numerical data from QNMs analysis in the right panel of Fig.\ref{kk} can be recast into the red dashed line to link the relations between $n$ and $m$. The blue solid line matches very well with the red dashed line as $m\lesssim1.5$. The quench rate in this figure is the same as in Fig.\ref{kk}, i.e., $\tau_Q=20$.}
\end{figure}
We count the vortex number at the final equilibrium state of the system, where the order parameter is well developed. From the above analysis, we expect that the vortex number $n$ would increase with the graviton mass,  and satisfying the following relation via Eq.\eqref{density} and Eq.\eqref{eq:xim}:
\begin{eqnarray}\label{eq:nm}
n\approx\frac{Ap}{\hat\xi^{2}}=\frac{Ap}{a^{2}}\left(1+b m^2\right)^2
\end{eqnarray}
The numerical results of the vortex number $n$ versus the graviton mass $m$ are shown as solid blue dots in Fig.\ref{nm} with error bars presenting the standard deviations. The solid blue line is the fitting relation of the numerical results as $n=17.82\times\left(1+0.1585 m^2\right)^2$ from Eq.\eqref{eq:nm}. Compared to the above $b\approx 0.1496$ in Eq.\eqref{eq:xim} from the right panel of Fig.\ref{kk}, we can see that the values of the coefficient before $m^2$ are close to each other within $5.6151\%$ errors. Estimates of parameter $b$ from these two methods, i.e., one is from the fitting of relation $n\sim m$, the other is from the fitting of $\hat\xi\sim m$, match each other very well! This also indicates that our conjecture of the relation between coherence length and graviton mass, viz. Eq.\eqref{eq:xim} is very trustable.

The overall factor in the front of Eq.\eqref{eq:nm} can even provide us the estimate of the probability $p$, which is the probability for the symmetry-breaking domains to form vortices, in Eq.\eqref{density}. We already knew $a\approx 0.7466$ in Eq.\eqref{eq:xim} from the right panel of Fig.\ref{kk}, and the size of the system is $A=50\times50=2500$, thus we can readily get $p\approx 17.82\times0.7466^2/2500\approx 0.003973$ and $n\approx Ap/\hat\xi^2\approx 9.9331\hat\xi^{-2}$. The later relation is plotted as red dashed line in the Fig.\ref{nm}, where $\hat\xi$ comes from the QNMs analysis in the right panel of Fig.\ref{kk}.  We see that the red dashed line fits very well with the blue fitting line except in the large regime of $m$, such as $m>1.5$.\footnote{In \cite{Blake:2013bqa,Davison:2013jba}, the authors stated that the relation $\tau_s^{-1}=\gamma m^2$ should hold as graviton mass is small, such that the momentum conservation is violated in a minor way. Therefore, if graviton mass is large enough, the above relation $\tau_s^{-1}=\gamma m^2$ may not hold again. Consequently, the analytic relation in Fig.\ref{nm} may not apply as $m$ is large enough.}
This discrepancy originates from the differences between the two fitted values of $b$, i.e, $b\approx 0.1496$ from Eq.\eqref{eq:xim} and $b\approx0.1585$ from Eq.\eqref{eq:nm}. However, the error between these two $b$'s is small, therefore, it is fair to say that our assumption of the relation between the coherence length and the graviton mass, viz. Eq.\eqref{eq:xim} is reliable.

\section{Conclusions and discussions}
\label{sec:conclusion}
In summary, we have investigated the effects of momentum dissipation induced by the graviton mass on the critical dynamics in the background of a $(3+1)$-dimensional dRGT massive gravity by taking advantage of the AdS/CFT correspondence. The KZM scaling laws of the vortex number and the ``freeze-out" time with respect to the quench rate were not affected by the graviton mass. This in turn indicated that the static and dynamic critical exponents were not changed by the existence of graviton mass. It further implied that the boundary field theory was always like a mean-field theory in spite of the graviton mass.

However, the outcome of vortex numbers depended on the graviton mass, since momentum dissipation would shrink the size of the symmetry-breaking domains. Inspired from Pippard's formula on the coherence length and the mean free path, we proposed an analytical relation between the coherence length and the graviton mass. This relation was verified from the QNMs analysis of the perturbations of the scalar field, which was a key result in our paper. We further investigated the relation between the number of vortices and the graviton mass both analytically and numerically, and the results were consistent with each other. Therefore, in conclusion, the momentum dissipation induced by graviton mass would increase the number of vortices, although it would not affect the KZM scaling laws.

\acknowledgments
We are grateful for Adolfo del Campo for helpful discussions and put forward a lot of valuable advices. Besides, we thank Qi-Rong Jiao, Hong-Da Lyu, Han-Qing Shi, Xin-Meng Wu, Chuan-Yin Xia and Jun-Kun Zhao for illuminating discussions. This work was supported by the National Natural Science Foundation of China (Grants No. 11675140, 11705005 and 11875095).

\appendix
\section{A Brief Review of dRGT massive Gravity}
\label{app}
Let us consider the action for an $(n+2)$-dimensional ghost-free dRGT massive gravity\cite{Vegh:2013sk,Cai:2014znn}
\begin{eqnarray}
  S =\frac{1}{16\pi G}\int d^{n+2} x\sqrt{-g} \left[R+\frac{n(n+1)}{L^2}-\frac{1}{4}F^2+m^2\sum_i^4c_i\mathcal{U}_i(g,\textbf f)\right],
\end{eqnarray}
where $L$ is the radius of the AdS$_{n+2}$ spacetime, $m$ stands for the graviton mass parameter, $\textbf f$ is a fixed symmetric tensor usually called the reference metric, $c_i$ are constants, and $\mathcal{U}_i$ are symmetric polynomials of the eigenvalues of the $(n+2)\times(n+2)$ matric $\mathcal{K}^\mu_{~\nu}\equiv\sqrt{g^{\mu\alpha}\textbf f_{\alpha\nu}}$:
\begin{eqnarray}
\mathcal{U}_1&=&[\mathcal{K}],\nonumber\\
\mathcal{U}_2&=&[\mathcal{K}]^2-[\mathcal{K}^2],\nonumber\\
\mathcal{U}_3&=&[\mathcal{K}]^3-3[\mathcal{K}][\mathcal{K}^2]+2[\mathcal{K}^3],\nonumber\\
\mathcal{U}_4&=&[\mathcal{K}]^4-6[\mathcal{K}^2][\mathcal{K}]^2+8[\mathcal{K}^3][\mathcal{K}]+3[\mathcal{K}^2]^2-6[\mathcal{K}^4].
\end{eqnarray}
The square root in $\mathcal{K}$ means $(\sqrt{A})^\mu_{~\nu}(\sqrt{A})^\nu_{~\lambda}=A^\mu_{~\lambda}$ and $[\mathcal{K}]=K^\mu_{~\mu}=\sqrt{g^{\mu\alpha}\textbf f_{\alpha\mu}}$. The equations of motion turn out to be
 \begin{eqnarray}
  R_{\mu\nu}-\frac{1}{2}Rg_{\mu\nu}-\frac{n(n+1)}{2L^2}g_{\mu\nu}-\frac12\left(F_{\mu\sigma}F_\nu^{\ \sigma}-\frac14g_{\mu\nu}F^2\right)+m^2\chi_{\mu\nu}&=&8\pi GT_{\mu\nu}, \\
  \nabla_\mu F^{\mu\nu}&=&0,
\end{eqnarray}
where
\begin{eqnarray}
\chi_{\mu\nu}&=&-\frac{c_1}{2}(\mathcal{U}_1g_{\mu\nu}-\mathcal{K}_{\mu\nu})-\frac{c_2}{2}(\mathcal{U}_2g_{\mu\nu}-2\mathcal{U}_1\mathcal{K}_{\mu\nu}
+2\mathcal{K}_{\mu\nu}^2)-\frac{c_3}{2}(\mathcal{U}_3g_{\mu\nu}-3\mathcal{U}_2\mathcal{K}_{\mu\nu}\nonumber\\
&&+6\mathcal{U}_1\mathcal{K}_{\mu\nu}^2-6\mathcal{K}_{\mu\nu}^3)-\frac{c_4}{2}(\mathcal{U}_4g_{\mu\nu}-4\mathcal{U}_3\mathcal{K}_{\mu\nu}
+12\mathcal{U}_2\mathcal{K}_{\mu\nu}^2-24\mathcal{U}_1\mathcal{K}_{\mu\nu}^3+24\mathcal{K}_{\mu\nu}^4)
\end{eqnarray}
One can choose a special form of reference metric as
\begin{eqnarray}
\mathbf f_{\mu\nu}={\rm diag}\{0,0,h_{ij}\}.
\end{eqnarray}
where $h_{ij}$ is the metric on a two-dimensional constant curvature space.
A black hole solution of (3+1)-dimensional dRGT massive gravity with this reference metric is \cite{Hu:2016mym}
 \begin{eqnarray}\label{eq:metric0}
  ds^2&=&-r^2f(r)dt^2+\frac{dr^2}{r^2f(r)}+r^2h_{ij}dx^idx^j,\label{eq:metric}\\
  f(r)&=&\frac{k}{r^2}+\frac{1}{L^2}-\frac{M}{r^3}+\frac{q^2}{4r^4}+\frac{c_1m^2}{2r}+\frac{c_2m^2}{r^2}\label{eq:metric1}
\end{eqnarray}
where $h_{ij}dx^idx^j$ is the line element for the 2-dimensional spherical, flat or hyperbolic space with  $k=\{1,0,-1\}$ respectively. $M$ is the black hole mass parameter while $q$ is the electric charge of it.
Hawking temperature of this black hole solution thus is
\begin{eqnarray}
T_{BH}=\frac{\left(r^2f(r)\right)'}{4\pi}\bigg|_{r=r_+}=\frac{1}{4\pi r_+}\left(k+\frac{3r_+^2}{L^2}-\frac{q^2}{4r_+^2}+c_1m^2r_++c_2m^2\right).
\end{eqnarray}
in which $r_+$ is the horizon of the black hole. In order to study the dynamics in this background, it is convenient to work under the Vaidya-AdS solution by making the transformation $d\upsilon=dt+dr/\left(r^2f(r)\right)$
\begin{eqnarray}
ds^2&=&-r^{2}f(r)d\upsilon^2+2d\upsilon dr+r^2h_{ij}dx^idx^j.
\end{eqnarray}
By setting
\be
\upsilon\to t,\ \ \ r\to 1/z,\ \ \ h_{ij}\to \delta_{ij},\ \ \ k\to0,\ \ \ q\to0,
\ee
one can obtain the line element \eqref{eq:1} in the main text.


\end{document}